\documentclass[showpacs,amsmath,amssymb,amsfonts,showkeys]{revtex4}

\usepackage{enumerate}
\usepackage{graphicx}
\usepackage{dcolumn}
\usepackage[usenames]{color}
\usepackage{ulem} 

\begin{document}
\draft
\date{\today}

\title{Reply to M. Campisi [arXiv: 1310.5556]}

\author{G. Baris Bagci}
\affiliation{ Department of Physics, Faculty of Science, Ege
University, 35100 Izmir, Turkey}

\author{Thomas Oikonomou}\thanks{Corresponding Author}
\email{thoik@physics.uoc.gr}

\affiliation{Department of Physics, University of Crete, 71003
Heraklion, (Hellas) Greece}

\affiliation{Department of Physics, Faculty of Science, Ege
University, 35100 Izmir, Turkey}

\pagenumbering{arabic}

\begin{abstract}
In response to M. Campisi's comment [arXiv: 1310.5556] on our
recent work [Phys. Rev. E 88, 042126 (2013)], we first point out
that the distribution used by Campisi is not the correct escort
distribution and further provide arguments showing that the
distributions obtained from the finite bath scenario are not
Tsallis distributions assuming the ergodicity of the total system.
We also comment on the role of evidence mentioned by M. Campisi.
\end{abstract}


\newpage \setcounter{page}{1}

\maketitle

\vskip1cm Before proceeding in our reply to the comment by M.
Campisi \cite{CampisiCOMMENT} on our recent work \cite{usmain}, we
first define the Tsallis entropy \cite{Tsallis1}

\begin{equation}\label{tsallis}
S_{q}=\frac{\int p^{q}d\Gamma -1}{1-q}
\end{equation}

\noindent without the multiplicative constant $k$, where $p$ is
the probability distribution, $q$ is the nonextensivity parameter,
and $\Gamma$ denotes the phase space variable. If one maximizes
the entropy above with the usual normalization condition $\int p
d\Gamma =1$ and the escort internal energy constraint $\frac{\int
p^{q}H_{S}d\Gamma }{\int p^{q}d\Gamma }=U$ where $H_S$ is the
system Hamiltonian and $U$ is the average energy, one obtains the
following Tsallis distribution

\begin{equation}\label{escort}
p = \left[ 1-\left( 1-q\right) \frac{\beta }{\int p^{q}d\Gamma
}\left( H_{S}-U\right) \right] ^{\frac{1}{1-q}}.
\end{equation}

\noindent where $\beta$ is the Lagrange multiplier associated with
the internal energy constraint \cite{Tsallis1}. The reader can
check this by inspecting Eqs. (3.197) and (3.198) in Ref.
\cite{Tsallis1}. Note that the normalization constant is
suppressed from here on, since it is irrelevant to the present
discussion.

Campisi, instead of Eq. \eqref{escort} above, prefers to raise the
exponent of the distribution in Eq. \eqref{escort} to the power of
$q$ and defining $P_{\text{Campisi}} = p^{q}$, writes

\begin{equation}\label{escort2}
P_{\text{Campisi}} = \Big[ 1-\left( 1-q\right) \beta \left(
H_{S}-U\right) \Big] ^{\frac{q}{1-q}}
\end{equation}

\noindent as can be seen from Eq. (4) and the footnote given under
the heading of Reference 6 in his comment. However, a quick
comparison of Eqs. \eqref{escort} and \eqref{escort2} reveals that
for $P_{\text{Campisi}} = p^{q}$ to be correct, one must also have
$\int p^{q}d\Gamma =1$ for $\frac{\beta }{\int p^{q}d\Gamma }$ to
be equal to $\beta$. If this assumption of Campisi i.e. $\int
p^{q}d\Gamma =1$ is however indeed realized, the equilibrium
Tsallis entropy, irrespective of the physical system under
scrutiny, always yields zero, which can be seen by substituting
$\int p^{q}d\Gamma =1$ into Eq. \eqref{tsallis}.

It is also important to note that even one grants Campisi the
condition $P_{\text{Campisi}} = p^{q}$ neglecting the above
discussion (as a result of using a different constraint in
obtaining $p$ for example), $P_{\text{Campisi}}$, being an
equilibrium distribution, should be normalized i.e., $\int
P_{\text{Campisi}}d\Gamma = 1$ i.e., $\int p^{q}d\Gamma =1$. This
normalization again, substituted in the Tsallis entropy given by
Eq. \eqref{tsallis}, leads to the same anomaly of assigning zero
entropy value for all systems at equilibrium which is not correct.

Concerning the equipartition theorem and finite baths, both us and
Campisi agree on the fact that the equipartition theorem (see Eq.
(5) in the comment by Campisi) should be intact in nonextensive
scenario i.e. the physical inverse temperature $\beta$ (and the
physical temperature itself for that matter) must be independent
of $q$. This is tantamount to the fact that the relevant term
appearing in the Tsallis distribution should simply be $\beta$ as
is also required by the finite bath scenario. The inspection of
the correct escort distribution in Eq. \eqref{escort} shows that
this is not so i.e. instead of $\beta$, we have a term
$\frac{\beta }{\int p^{q}d\Gamma }$ which is $q$-dependent. We
might choose to consider this term as the physical inverse
temperature, but this happens at the expense of violating the
equipartition theorem (which simply states that the temperature
must be $q$-independent) as we noted in our work \cite{usmain}.
Other option is to set $\int p^{q}d\Gamma =1$ in Eq.
\eqref{escort}. But, as we explained above, the Tsallis entropy
given by Eq. \eqref{tsallis} then becomes zero for any system at
equilibrium which is nonsensical.

When a physical system is in weak contact with a thermal bath of
finite constant heat capacity $C_B$, the probability distribution
of the system reads

\begin{equation}\label{positive}
\rho=\left( E_{tot}-H_{S}\right) ^{C_{B}-1}\,, 
\qquad C_B > 0\,,
\end{equation}

\noindent where $E_{tot}$ is the total energy of the system and
the bath. When the bath has negative finite heat capacity i.e.
$C_B < 0$, albeit constant again, the system distribution
possesses the same form, but the term $\left(
E_{tot}-H_{S}\right)$ is replaced by $\left( H_{S}-E_{tot}\right)$
as Campisi notes in Eqs. (1) and (2) in his comment. To be able to
produce the Tsallis distributions from these system distributions,
as Campisi notes (see below Eq. (3) in his comment), one has to
use the relation

\begin{equation}\label{relation}
C_B = \frac{1}{1-q}
\end{equation}

\noindent for all values of $q$. The system probability
distributions given by Eq. \eqref{positive} (and its counterpart
for the case $C_B < 0$) have to satisfy two important physical
limits, namely, the limits $ C_{B}\rightarrow \infty$ and $
C_{B}\rightarrow 0$. The former limit corresponds to the
well-known canonical distribution when the heat capacity of the
bath becomes infinite whereas the latter yields the microcanonical
ensemble where the system is completely isolated. These limits are
also mathematically important, since the limit $C_{B}\rightarrow
0$ for example must exist due to the continuity of the finite
baths formalism. Note that both these limits are also agreed on by
Campisi \cite{CampisiEPL}. These limits can easily be taken to
produce the canonical distribution and the microcanonical Dirac
delta as can be easily verified by using Eq. \eqref{positive} and
its $C_B < 0$ counterpart.

On the other hand, if the relation in Eq. \eqref{relation} should
make any sense within the Tsallis formalism, these same limits
must be taken in terms of the $q$ values in accordance with Eq.
\eqref{relation}. The limiting behavior $C_{B}\rightarrow \infty$,
due to Eq. \eqref{relation}, corresponds to the limit
$q\rightarrow 1$ of the escort distribution in Eq. \eqref{escort},
yielding indeed to the canonical distribution in nonextensive
thermostatistics.

However, the second limiting behavior $C_{B}\rightarrow 0$
corresponding to the microcanonical ensemble is problematic for
the Tsallis distributions in Eq. \eqref{escort} although this
limit can easily be taken to obtain the Dirac delta distribution
through the finite bath system distributions given by Eq.
\eqref{positive} and its counterpart for $C_B < 0$. The inspection
of Eq. \eqref{relation} shows that this limiting behavior
corresponds to the limit $q \rightarrow \pm \infty$, depending on
whether one approaches from the right $C_{B}\rightarrow 0^{+}$
(i.e. $q\rightarrow -\infty$) or left $C_{B}\rightarrow 0^{-}$
(i.e. $q\rightarrow +\infty$). The limits $q \rightarrow \pm
\infty$ cannot be realized in the Tsallis escort distributions
given by Eq. \eqref{escort}: The first limit i.e. $q \rightarrow
-\infty$ cannot be realized, since the Tsallis entropy given by
Eq. \eqref{tsallis} is concave only for $q>0$, which can easily be
seen from the condition $\frac{\partial ^{2}S_{q}}{\partial
p^{2}}<0$ i.e.

\begin{equation}\label{concavity}
\frac{\partial ^{2}S_{q}}{\partial p^{2}}<0
\qquad\Longrightarrow\qquad
\frac{q\left( q-1\right) p^{q-2}}{\left( 1-q\right)
}=-qp^{q-2}<0
\qquad\Longrightarrow\qquad
 q > 0.
\end{equation}

\noindent This implies that even if one mathematically forces the
limit $q \rightarrow -\infty$ on the Tsallis distributions and
obtains a microcanonical distribution, one will never know that
this is the distribution maximizing the Tsallis entropy.

Concerning the limit $q \rightarrow +\infty$, the permissible
$q$-values are limited within the range $1<q<1+\frac{2}{d N_S}$
where $N_S$ and $d$ denote the number of particles in the system
and the dimensionality, respectively as Lutsko and Boon
\cite{LB2011} first showed, and as we confirmed (see the paragraph
below Eq. (17) in our work \cite{usmain}). Due to this permissible
interval of $q$ values, the limit $q \rightarrow +\infty$ requires
either $d$, or the number of particles in the system $N_S$ to attain
a value of zero which is nonsensical whereas the limit
$C_B\rightarrow 0$, using the system distribution in Eq.
\eqref{positive} and its $C_B < 0$ counterpart without relating it
to the parameter $q$, is permissible and unproblematic. We also
note that the above argument is applicable to any kind of Tsallis
distributions, independently of the constraints, since concavity
is a property of the Tsallis entropy itself, and the Tsallis
distributions cannot attain the limit $q \rightarrow +\infty$ due
to the normalization requirement.

The simulation Campisi mentions in the fourth paragraph of his
comment is not in contradiction with our results, but instead
supports them: since there is neither a single occurrence of the
nonextensivity parameter $q$ nor Tsallis distributions themselves
in the aforementioned simulations and related calculations in Ref.
\cite{CampisiEPL}, the work by Campisi \textit{et al}.
\cite{CampisiEPL} simply shows that the finite bath distributions
can be studied in its full generality without any additional
nonextensive ingredient.

%

In sum, the equilibrium distributions originating from the finite
baths, assuming the ergodicity of the total compound, are not
Tsallis distributions. However, our results do not exclude the
possibility of Tsallis distributions in non-ergodic systems.


\begin{thebibliography}{00}

\bibitem{CampisiCOMMENT} M. Campisi, arXiv: 1310.5556.

\bibitem{usmain} G. B. Bagci and T. Oikonomou, Phys. Rev. E \textbf{88}, 042126 (2013).

\bibitem{Tsallis1} C. Tsallis, \textit{Introduction to Nonextensive Statistical Mechanics},
(Springer, New York) 2009.

\bibitem{LB2011} J. F. Lutsko and J. P. Boon, Europhys. Lett. \textbf{95}, 20006 (2011).

\bibitem{CampisiEPL} M.Campisi, F. Zhan and P. H\"{a}nggi, EPL \textbf{99} (2012) 60004.






























\end{thebibliography}
\end{document}